\documentstyle{mn}
\def\spose#1{\hbox to 0pt{#1\hss}}
\def\simlt{\mathrel{\spose{\lower 3pt\hbox{$\mathchar"218$}}
     \raise 2.0pt\hbox{$\mathchar"13C$}}}
\def\simgt{\mathrel{\spose{\lower 3pt\hbox{$\mathchar"218$}}
     \raise 2.0pt\hbox{$\mathchar"13E$}}}

\def\OIII{[O{\footnotesize~III}]}
\newcommand{\cntr}[2]{\multicolumn{#1}{c}{#2}}

\begin{document}
\title[1--5\,$\mu$m imaging of 3CRR galaxies]{1--5\,$\mu$m imaging
of 3CRR galaxies: The $K$--$z$ relation and the geometry of the torus}
\author[C. Simpson et al.]{Chris Simpson,$^1$ Martin Ward,$^2$ and
J. V. Wall$^3$\\
$^1$Subaru Telescope, National Astronomical Observatory of Japan, 650
N. A`oh\={o}k\={u} Place, Hilo, HI 96720, U.S.A.\\
$^2$X-ray Astronomy Group, Department of Physics and Astronomy,
University of Leicester, Leicester LE1 7RH\\
$^3$Astrophysics, Department of Physics, University of Oxford, Oxford
OX1 3RH}

\maketitle

\begin{abstract}
It has been claimed by Taylor et al.\ that the low-redshift end of the
$K$--$z$ relation for radio galaxies is too bright by about half a
magnitude due to contributions from the obscured quasar nuclei. Such
a result has major implications for the use of the $K$-band Hubble
diagram in understanding the cosmological evolution of radio galaxies.
In this paper we present 1--5\,$\mu$m imaging data of a
nearly-complete sample of low-redshift radio galaxies; this approach
allows us to accurately determine the strengths of any unresolved
nuclear components in the galaxies. We detect nuclear sources in five
targets, whose broad-band colours are consistent with reddened quasar
spectra. In all five cases the ratio of the inferred intrinsic
near-infrared luminosity to the narrow-line luminosity is typical of
quasars. We find a correlation between the inferred nuclear extinction
and core-to-lobe ratio, which places constraints on the geometry of
the torus. We find evidence for a shift of the $K$--$z$ relation to
fainter magnitudes, but by a much smaller amount ($\sim 0.1$\,mag)
than Taylor et al.\ determined. Under the assumption that the nuclear
sources in radio galaxies have the same intrinsic near-infrared
spectra as quasars, our multi-wavelength images allow us to limit any
possible shift to less than 0.3\,magnitudes.
\end{abstract}

\section{Introduction}

It has been proposed that powerful FR\,II (see Fanaroff \& Riley 1974)
radio galaxies and radio loud quasars are intrinsically the same
objects, with all observed differences being a result of the angle
between the line of sight and the radio axis (Scheuer 1987; Barthel
1989). While objects classified as ``narrow-line radio galaxies'' do
not display obvious broad lines in their optical spectra, broad lines
have been seen either in the near-infrared (e.g.\ Hill, Goodrich \&
DePoy 1996) or in polarized light (e.g.\ Young et al.\ 1996). In
addition, the detection of unresolved nuclei in near-infrared
observations of radio galaxies (e.g.\ Djorgovski et al.\ 1991;
Simpson, Ward \& Wilson 1995), where the optical depth is lower,
indicates the presence of a quasar-like nucleus in the centres of
these objects. Since quasars can outshine their host galaxies by
factors ranging from a few (e.g.\ Dunlop et al.\ 1993; Taylor et al.\
1996, hereafter T96) to more than 100 in extreme cases (e.g.\ PDS~456;
Simpson et al.\ 1999), even if only a few per cent of the nuclear
light is transmitted it can still make a major contribution to the
integrated luminosity.

It has long been known that the brightest radio galaxies, i.e.\ those
in the 3CRR catalogue of Laing, Riley \& Longair (1983), are brighter
in the near-infrared by about a magnitude at $z \approx 1$ than they
are at low redshift. This amount of dimming as cosmic time increases
is consistent with passive evolution of their stellar populations
(Lilly \& Longair 1984; Lilly, Longair \& Allington-Smith 1985b).
While some 3C radio galaxies at $z \approx 1$ suffer significant
contamination from non-stellar radiation (Rawlings et al.\ 1995;
Simpson, Rawlings \& Lacy 1999), this does not have a significant
impact on the locus of the $K$--$z$ relation. However, radio galaxies
from the fainter 6C and B2 radio surveys are 0.6\,mag fainter at $K$
than the 3C objects at these redshifts (Eales et al.\ 1997), whereas
there is no correlation between radio and host galaxy luminosities
locally. Furthermore, the 6C/B2 galaxies lie {\em below\/} the
no-evolution curve.

The possibility that there is significant non-stellar contamination at
{\em low\/} redshift is one that therefore needs to be investigated. A
study by T96 claimed that the true locus of the $K$--$z$ relation at
low redshift may be half a magnitude fainter than previously thought
due to contamination from the non-stellar nucleus. In this paper, we
perform a similar analysis to that of T96 on an unbiased sample of
radio galaxies from the catalogue of Laing et al.\ (1983). In
Section~2 we describe our sample selection and observational
technique. In Section~3 we briefly describe the two-dimensional
fitting procedure used to separate the nuclear and host galaxy
properties in our $J$ and $K$ images, and present the results of this
analysis, together with aperture photometry of the longer-wavelength
images. In Section~4 we use these results to examine the nature of the
stellar and non-stellar components of radio galaxies, and infer the
intrinsic luminosities of the nuclear sources and the line of sight
obscurations towards them. In Section~5 we attempt to infer the
geometry of this obscuring material, and in Section~6 we discuss the
implications of our analysis on the low-redshift end of the $K$--$z$
relation and compare our results with those of T96. Finally, we
summarize our results in Section~7.  Throughout this paper we adopt
$H_0 = 50$\,km\,s$^{-1}$\,Mpc$^{-1}$, $q_0 = 0.5$, and $\Lambda = 0$.
Our convention for spectral index, $\alpha$, is such that the flux
density at a frequency $\nu$, $S_\nu \propto \nu^{-\alpha}$.

\section{Sample Selection and Observations}

We selected our targets from the 3CRR catalogue of Laing et al.\
(1983). This catalogue is essentially complete and, by virtue of being
selected at low frequency, does not preferentially include sources
observed close to the radio axis whose fluxes are enhanced by Doppler
boosted cores. It is therefore an unbiased sample of radio galaxies,
except in the sense that it consists of the most radio-luminous
objects. We excluded objects with known broad H$\alpha$, as determined
by the high-quality spectrophotometric observations of Laing et al.\
(1994, hereafter L94); we did not however exclude 3C~234 since the
broad H$\alpha$ seen is believed to be entirely scattered (Antonucci
1984; Young et al.\ 1998). This criterion supplied an upper redshift
cutoff of $z<0.43$, in order that H$\alpha$ still be present in the
optical spectra. From those objects visible from the 3.8-m United
Kingdom Infrared Telescope (UKIRT) in early January, we selected only
those galaxies with strong optical emission lines (Class~A; Hine \&
Longair 1979). The reasons for this were twofold. First, the
low-ionization (``optically dull'', Class~B) radio galaxies may not be
part of the quasar--radio galaxy unification paradigm -- they are
predominantly low-power objects with an FR\,I radio
morphology. Secondly, since emission-line luminosity is a good
indicator of the power of the active nucleus (Rawlings \& Saunders
1991), it follows that these objects will have less luminous nuclei in
the near-infrared and are therefore likely to be below our detection
threshold. We were able to observe ten of the eleven objects thus
selected, with 3C~123 being excluded from our programme solely because
it was never at a suitable hour angle when the time came to choose the
next target. We present details of the sample in
Table~\ref{tab:sample}, including observed 178\,MHz and
\OIII~$\lambda$5007 fluxes and rest frame luminosities.  Although the
data of L94 is of a consistently high quality, their choice of a
narrow slit necessarily meant a loss of extended emission line
flux. This effect is especially pronounced in the closest radio
galaxies, where the line emission can be extended over many arcseconds
(see, e.g., Baum et al.\ 1988). We have therefore chosen not to use
fluxes from L94's data in those cases where a larger aperture
measurement is available that contains a significantly larger flux. We
were unable to locate large aperture measurements for 3C~42 or 3C~153,
although these are the two most distant objects in our sample and we
can expect the spectroscopic slit of L94 to include virtually all the
line emission.

\begin{table*}
\caption[]{The sample of radio galaxies studied. Radio fluxes and
spectral indices are from Laing et al.\ (1983), although the radio
fluxes have been multiplied by 1.09 to put them on the scale of Baars
et al.\ (1977). The Galactic extinctions used to deredden the \OIII\
luminosities are from NED. References for \OIII\ flux: YO = Yee \& Oke
(1978), assuming $\lambda5007/\lambda4959=3$; L = L94 and Laing et
al.\ (2000).}
\label{tab:sample}
\begin{center}
\begin{tabular}{lcccccr@{}lcccc}
\hline
Name & IAU name & $z$ & $S_{\rm178\,MHz}$ & $\alpha$ & $\log
L_{\rm178\,MHz}$ & \cntr{2}{$f_{5007}$} & \cntr{2}{$\log L_{5007}$ (W)} &
\OIII\ & $E(B-V)$ \\
& & & (Jy) & & (W\,Hz$^{-1}$) & \cntr{2}{($10^{-18}$\,W\,m$^{-2}$)} &
observed & dereddened & ref & \\
\hline
3C~33  & 0106+130 & 0.0595 & 59.3 & 0.76 & 26.95 & \hspace*{5mm} 108&   &
35.20 & 35.23 & YO & 0.02 \\
3C~42  & 0125+288 & 0.3950 & 13.1 & 0.73 & 27.91 & \hspace*{5mm}   1&.6 &
34.96 & 35.01 & L  & 0.05 \\
3C~79  & 0307+169 & 0.2559 & 33.2 & 0.92 & 27.93 & \hspace*{5mm}  48&.5 &
36.08 & 36.19 & L  & 0.10 \\
3C~84  & 0316+413 & 0.0172 & 66.8 & 0.78 & 25.93 & \hspace*{5mm} 544&   &
34.84 & 35.07 & L  & 0.17 \\
3C~98  & 0356+102 & 0.0306 & 51.4 & 0.78 & 26.32 & \hspace*{5mm}  76&   &
34.48 & 34.65 & YO & 0.13 \\
3C~153 & 0605+480 & 0.2769 & 16.7 & 0.66 & 27.73 & \hspace*{5mm}   2&.2 &
34.80 & 35.04 & L  & 0.23 \\
3C~171 & 0651+542 & 0.2384 & 21.3 & 0.87 & 27.68 & \hspace*{5mm}  21&.7 &
35.68 & 35.74 & L  & 0.06 \\
3C~192 & 0802+243 & 0.0598 & 23.0 & 0.79 & 26.54 & \hspace*{5mm}  57&   &
34.94 & 34.98 & YO & 0.03 \\
3C~223 & 0936+361 & 0.1368 & 16.0 & 0.74 & 27.10 & \hspace*{5mm}  43&.8 &
35.53 & 35.53 & L  & 0.00 \\
3C~234 & 0958+290 & 0.1848 & 34.2 & 0.86 & 27.68 & \hspace*{5mm} 208&   &
36.45 & 36.47 & YO & 0.01 \\
\hline
\end{tabular}
\end{center}
\end{table*}

All the images presented here were obtained on UKIRT in photometric
conditions, using IRCAM3 with a nominal scale of
0.286\,arcsec\,pixel$^{-1}$. Most observations were made on the nights
of UT 1996 Jan 3--6, although 3C~33, 3C~153, and 3C~171 were observed
during an earlier run on UT 1995 Jan 3--5 as part of a pilot
study. Each target was observed through the {\it JKL$'$M\/} filters,
with typical on-source exposure times of 18\,minutes at $J$ and $K$
and 30\,minutes at $L'$ and $M$. Most of the radio galaxies are
sufficiently compact at $J$ and $K$ compared to the size of the array
that it was possible to use standard jittering techniques and
therefore there was no need for ``off-source'' observations to produce
a flatfield image. In these cases, each observation consisted of
taking two sets of nine jittered frames of 60\,s duration each (split
into shorter, background-limited exposures). For those objects with $z
< 0.1$, however, nearby regions of sky were used to enable
flatfielding and sky subtraction. The much higher background in the
thermal-infrared ($L'$ and $M$) meant that only the central regions of
the radio galaxies were detected with any significance, and even the
closest objects could be jittered about the array. However, to permit
accurate flatfielding given the rapid background variations, shorter
exposure times were necessary and each observation was split into sets
of five frames of 30\,s ($L'$) or 15\,s ($M$), each of which was again
comprised of a number of shorter, background-limited exposures. Due to
time constraints, we were unable to observe two of our targets (3C~42
and 3C~84) at $M$.

Data reduction was performed in the following manner. Each observation
was split into its sets of nine (at $J$ and $K$) or five (at $L'$ and
$M$) separate exposures, and these were dealt with as separate
entities.  First, an appropriate dark frame was subtracted from all
the images. Then a flat field was produced by median-filtering either
the sky frames (if the source was at $z < 0.1$) or the object frames
(for cases where no off-source observations were made), after scaling
each frame to have the same median pixel value. Each object frame was
then divided by a normalized version of this flatfield, and these
images were registered, using stars and galaxies visible on the
individual images where possible, or the nominal telescope offsets
otherwise. An iterative technique was used to determine the DC offsets
between regions of overlap in separate frames, and the images were
averaged. At this stage, the separate sets were averaged together to
produce the final image. The seeing was measured from stars in the
individual exposures to be in the range 1.0--1.2\,arcsec FWHM (about 4
pixels) and the registration uncertainties did not cause the FWHM of
stars in the final coadded images to increase by a measurable amount.

\begin{table*}
\caption[thermphot]{Photometry in a 3-arcsec aperture. Limits are
$3\sigma$.}
\label{tab:phot}
\centering
\begin{tabular}{lccr@{ }c@{ }lr@{ }c@{ }l}
Galaxy & $J$ & $K$ & \cntr{3}{$L'$} & \cntr{3}{$M$} \\
\hline
3C~33  & $14.41\pm0.04$ & $13.16\pm0.04$ &
11.89&$\pm$&0.08 & 11.17&$\pm$&0.13 \\
3C~42  & $17.52\pm0.02$ & $16.06\pm0.03$ &
$>15.08$&& & \cntr{3}{not observed} \\
3C~79  & $16.37\pm0.03$ & $14.79\pm0.06$ &
12.96&$\pm$&0.08 &$>12.43$&& \\
3C~84  & $13.41\pm0.03$ & $11.90\pm0.05$ &
9.66&$\pm$&0.06 & \cntr{3}{not observed} \\
3C~98  & $13.83\pm0.02$ & $12.60\pm0.03$ &
12.19&$\pm$&0.07 & $>12.36$&& \\
3C~153 & $16.71\pm0.04$ & $15.25\pm0.04$ &
$>14.97$&& & $>12.71$&& \\
3C~171 & $16.97\pm0.04$ & $15.71\pm0.04$ &
$>15.23$&& & $>12.44$&& \\
3C~192 & $14.43\pm0.03$ & $13.46\pm0.05$ &
13.39&$\pm$&0.08 & $>12.28$&& \\
3C~223 & $16.02\pm0.04$ & $14.58\pm0.05$ &
12.91&$\pm$&0.07 & 11.79&$\pm$&0.20 \\
3C~234 & $15.33\pm0.03$ & $13.08\pm0.05$ &
10.50&$\pm$&0.06 & 9.91&$\pm$&0.08 \\
\hline
\end{tabular}
\end{table*}

Images of photometric standard stars were taken throughout the course
of each night. One standard star was imaged five times in each filter
for each radio galaxy. Multiple coadds were used both to improve the
signal-to-noise ratio, and to ensure that each observation had a
similar exposure time to the individual radio galaxy observations. In
this way, the point spread function (PSF) of the standard star would
sample the same longer-term seeing variations as the galaxy images.
Flux calibration solutions were determined separately for each night
of observation, with the r.m.s.\ dispersion for each night typically
being 4\% at $J$ and $K$ and 7\% at $L'$ and $M$. Aperture photometry
from our images agrees with that of Lilly \& Longair (1984) and Lilly,
Longair \& Miller (1985a) to within the quoted uncertainties. In
Table~\ref{tab:phot} we list the photometry measured in a 3-arcsec
aperture, which is more appropriate for detecting red nuclear sources.

\section{Separation of nucleus and host galaxy}
\label{sec:nuclei}

Thermal emission from the telescope is very strong longward of
2.5\,$\mu$m, resulting in a vastly increased background and a large
drop in sensitivity. In addition, the strength of the quasar nucleus
relative to the host galaxy increases with wavelength due to its red
colour. We therefore undertake different analyses for the short ({\it
JK\/}) and long ({\it L$'$M\/}) wavelength data.

In our $J$ and $K$ images, the host galaxies are well-detected out to
large radii and are likely to dominate over the nuclei, even in fairly
small apertures. Methods which attempt to estimate the flux of a
nuclear source by using aperture photometry to correct small aperture
measurements for starlight are prone to overestimate the strength of
the source (Simpson 1994b) and it is necessary to model the galaxy in
order to obtain a reliable measurement. At longer wavelengths, the
host galaxy is not detected with great significance due to the bright
thermal background, and simpler methods can be employed.

\subsection{Two-dimensional modelling}
\label{sec:jk}

To enable the most direct comparison possible of our results with
those of T96, we have used an almost identical analysis technique on
our data. We refer the reader to that paper for a more detailed
description of the method, and for a discussion of its advantages over
the more common (and simpler) one-dimensional analysis (e.g.\ Simpson
et al.\ 1995).

The two-dimensional approach involves constructing a model image
derived from fitting parameters and comparing this with the actual
data, using an error frame to allow a quantitative $\chi^2$
minimization. The assumed model was an elliptical galaxy obeying a de
Vaucouleurs law (de Vaucouleurs 1948) and an unresolved nuclear
source. There are five parameters used to construct the
model:
\begin{enumerate}
\item The luminosity of the nuclear source, $L_{\rm nuc}$
\item The luminosity of the host galaxy, $L_{\rm host}$
\item The effective (half-light) radius of the host galaxy, $r_{\rm e}$
\item The position angle of the major axis of the host galaxy, $\Theta$
\item The axial ratio of the host galaxy, $a/b$
\end{enumerate}

The pixel location of the centre of the galaxy, ($x_{\rm c}$,$y_{\rm
c}$), was determined using iterative centroiding, and construction of
the model galaxy proceeded as follows. Each pixel in the image was
divided into 400 subpixels, and the flux in each of these subpixels
was computed based on the parameters of the galaxy model. This amount
of subpixellation ensures that the flux in the central pixel is never
underestimated by more than 0.14\%, and any overestimation of the
nuclear source flux as a result will be negligible. The nuclear
source was then added by increasing the value of the central pixel
($x_{\rm c}$,$y_{\rm c}$) by the appropriate amount. This
nucleus-plus-galaxy model was convolved with the adopted PSF to
produce a simulated image.

The quality of fit for the simulated image was determined by the
$\chi^2$ statistic,
\[
\chi^2 = \sum \left( \frac{z(x,y) - \zeta(x,y)}{\sigma(x,y)} \right) ^2 ,
\]
where $z(x,y)$ and $\zeta(x,y)$ are the values of the ($x$,$y$) pixel
in the real data and the simulated image respectively, and
$\sigma(x,y)$ is the uncertainty assigned to this pixel. We
constructed an error frame in exactly the same manner as T96, although
the sampling errors are smaller for our data due to the smaller pixel
scale. This procedure results in good error estimation, and the images
of the weighted residuals from the best-fit models appeared
structureless. Companion objects and foreground stars were effectively
excluded from the fitting procedure at this point by setting the pixel
values of the error frame to some very large value in regions where a
contaminating source was visible, thus giving such pixels a negligibly
small weight in determining the quality of fit.

In virtually all cases, the radio galaxy occupied only a fraction of
the total frame. To speed up the computational process, we therefore
used only a subregion of the image for our analysis. We selected a
square region which just encompassed the isophote whose intensity was
0.01\% of the sky brightness, the level at which flatfielding
uncertainties are expected to become important. Typically such a
region was about 30\,arcsec on a side. The exception was 3C~84, whose
proximity meant that such an isophote was larger than the entire
IRCAM3 field of view, and so we used the full array. We deviate from
the method of T96 in the optimization algorithm we use. Instead of the
Polak-Ribiere minimization of T96, we used a simulated annealing
algorithm (Press et al.\ 1996). We first tested the fitting procedure
on simulated galaxies to which noise had been added. After confirming
that it successfully recovered the input parameters of the model, the
fitting procedure was performed ten times in $J$ and $K$ for each
radio galaxy, with two different starting simplexes for each of the
five different PSFs. The results of the fit which produced the lowest
value of $\chi^2$ were used in later analysis.

We estimate the uncertainties in our fitting parameters by considering
the variations in the results from the ten separate fits. However, not
all PSFs were able to produce good fits, by which we mean a value of
$\chi^2$ close to that of the best fit, and we rejected those which
did not. As expected, only the nuclear luminosity depends strongly on
the PSF. Typically (for galaxies with $a/b \simgt 1.1$), the position
angle is determined to 5\degr\ or better, while the uncertainty is as
large as 10\degr\ for 3C~192 (with $a/b \approx 1.03$, the roundest
galaxy in our sample). The axial ratio is always determined to better
than 5\%. Although the host galaxy luminosity is generally determined
to $\sim 5$\% r.m.s., the uncertainty in the effective radius is
larger, ($\sim 10$\%), and the peak surface brightness can be
uncertain by as much as $\sim 20$\% (with there being a strong
anti-correlation between effective radius and peak surface
brightness). The fits we performed in the $J$ and $K$ filters for each
galaxy were completely independent. However, it is expected that some
quantities, such as the axial ratio and the position angle of the
major axis, should be very similar in the two bands. The difference
between the values of these quantities derived for each filter can
therefore also be taken as a measure of the uncertainty.

\begin{figure*}
\includegraphics{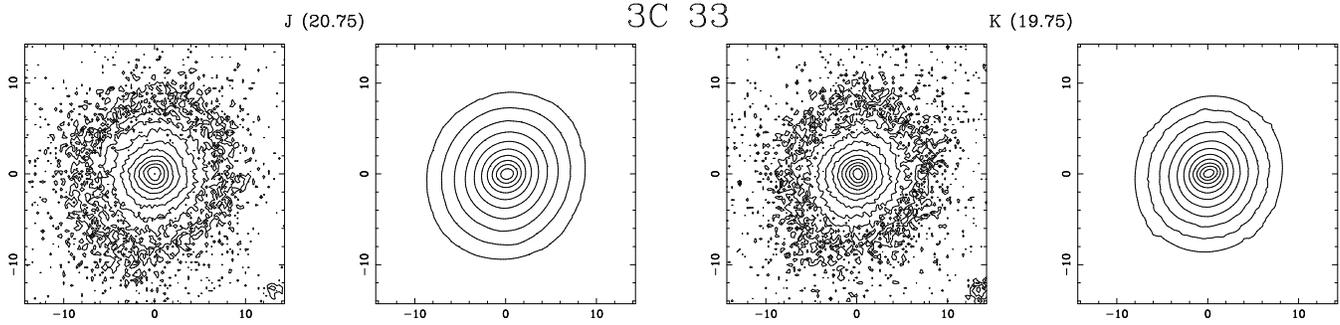}
\vspace*{48mm}
\caption[]{Images of 3C~33 and the best-fitting model at $J$ (left two
panels) and $K$ (right two panels). The number in parenthesis
indicates the value of the lowest contour plotted (in
mag\,arcsec$^{-2}$), with successive contours spaced at 0.5\,mag
intervals. North is up and east is to the left. The axes are labelled
in arcseconds. These images have not been smoothed.}
\label{fig:cont33}
\end{figure*}

\begin{figure*}
\includegraphics{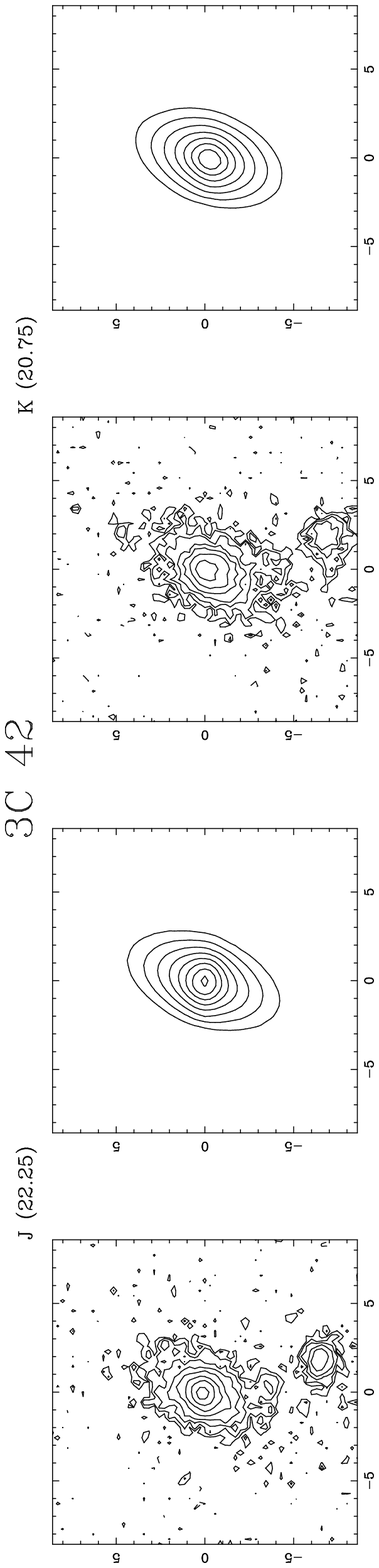}
\vspace*{48mm}
\caption[]{As Fig.~\ref{fig:cont33}, but for 3C~42.}
\label{fig:cont42}
\end{figure*}

\begin{figure*}
\includegraphics{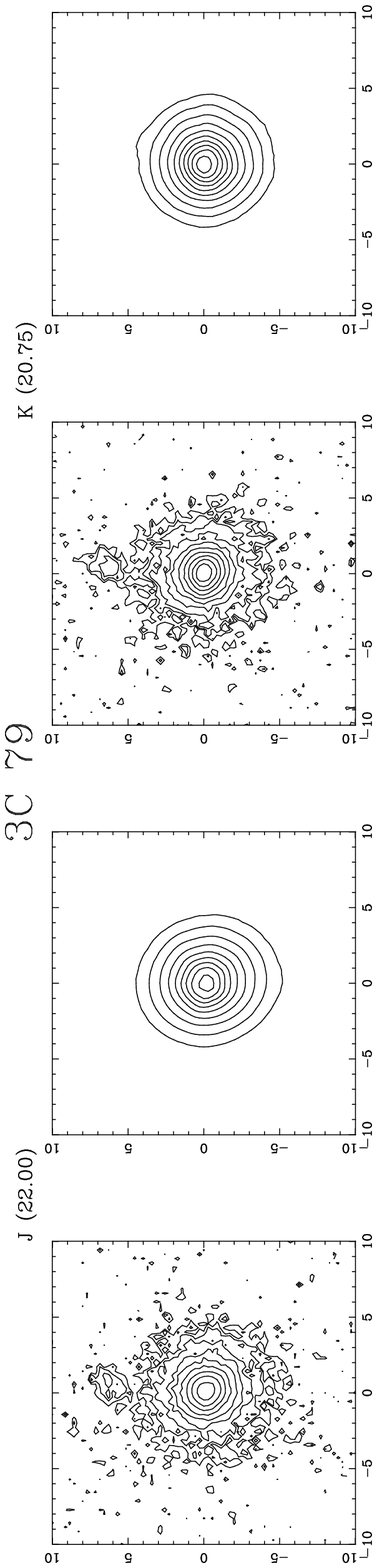}
\vspace*{48mm}
\caption[]{As Fig.~\ref{fig:cont33}, but for 3C~79.}
\label{fig:cont79}
\end{figure*}

\begin{figure*}
\includegraphics{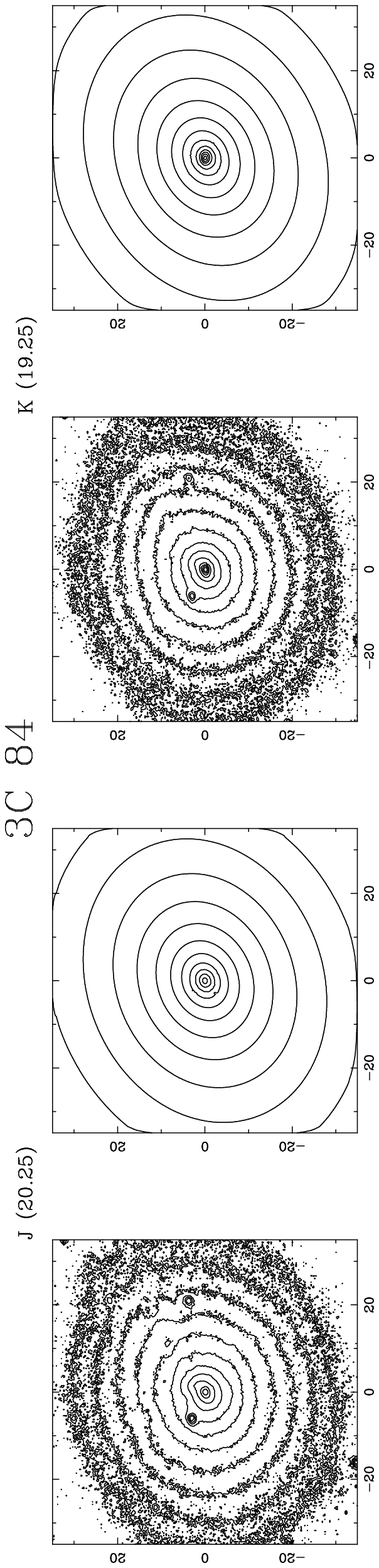}
\vspace*{48mm}
\caption[]{As Fig.~\ref{fig:cont33}, but for 3C~84.}
\label{fig:cont84}
\end{figure*}

\begin{figure*}
\includegraphics{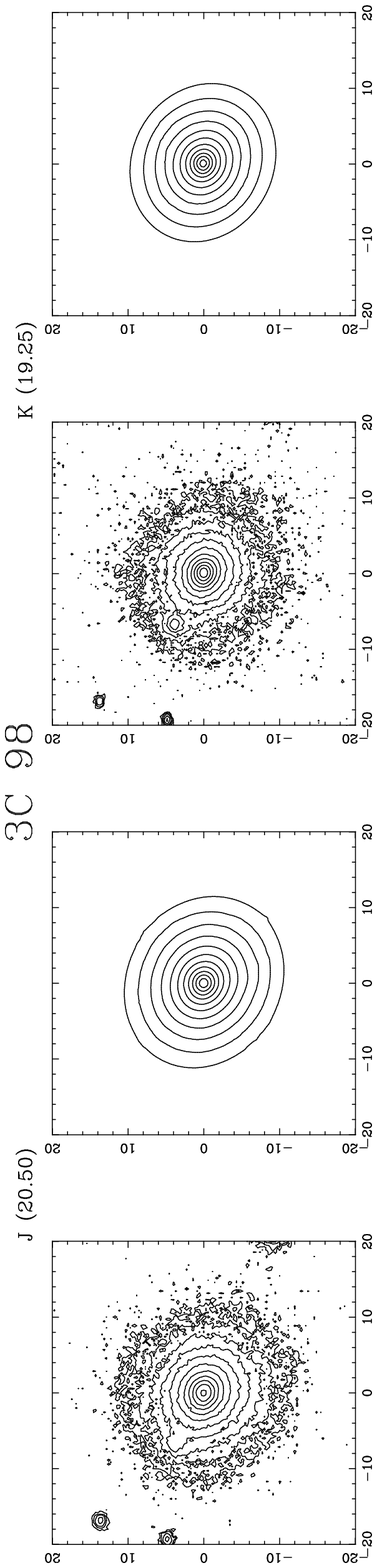}
\vspace*{48mm}
\caption[]{As Fig.~\ref{fig:cont33}, but for 3C~98.}
\label{fig:cont98}
\end{figure*}

\begin{figure*}
\includegraphics{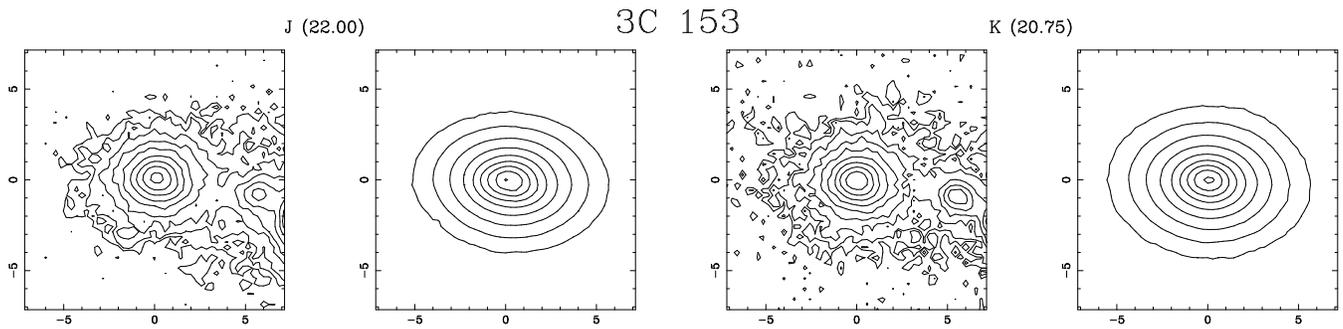}
\vspace*{48mm}
\caption[]{As Fig.~\ref{fig:cont33}, but for 3C~153.}
\label{fig:cont153}
\end{figure*}

\begin{figure*}
\includegraphics{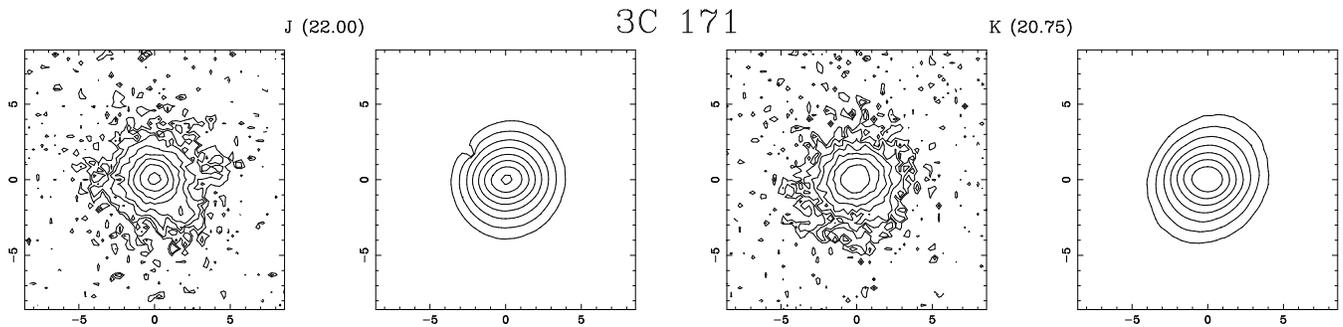}
\vspace*{48mm}
\caption[]{As Fig.~\ref{fig:cont33}, but for 3C~171.}
\label{fig:cont171}
\end{figure*}

\begin{figure*}
\includegraphics{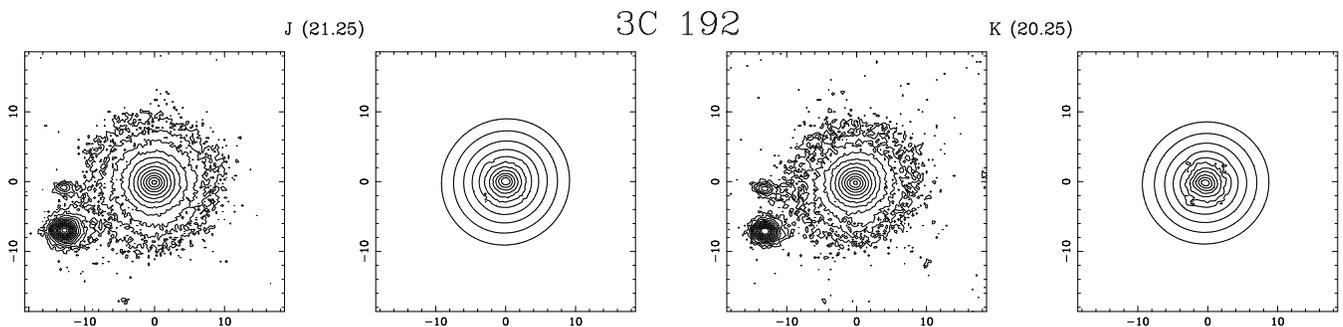}
\vspace*{48mm}
\caption[]{As Fig.~\ref{fig:cont33}, but for 3C~192.}
\label{fig:cont192}
\end{figure*}

\begin{figure*}
\includegraphics{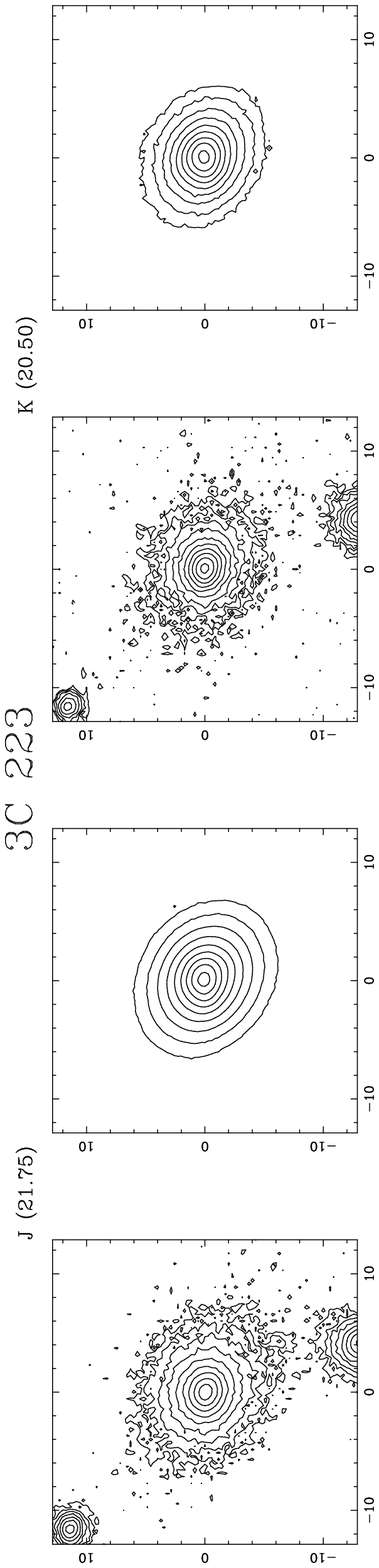}
\vspace*{48mm}
\caption[]{As Fig.~\ref{fig:cont33}, but for 3C~223.}
\label{fig:cont223}
\end{figure*}

\begin{figure*}
\includegraphics{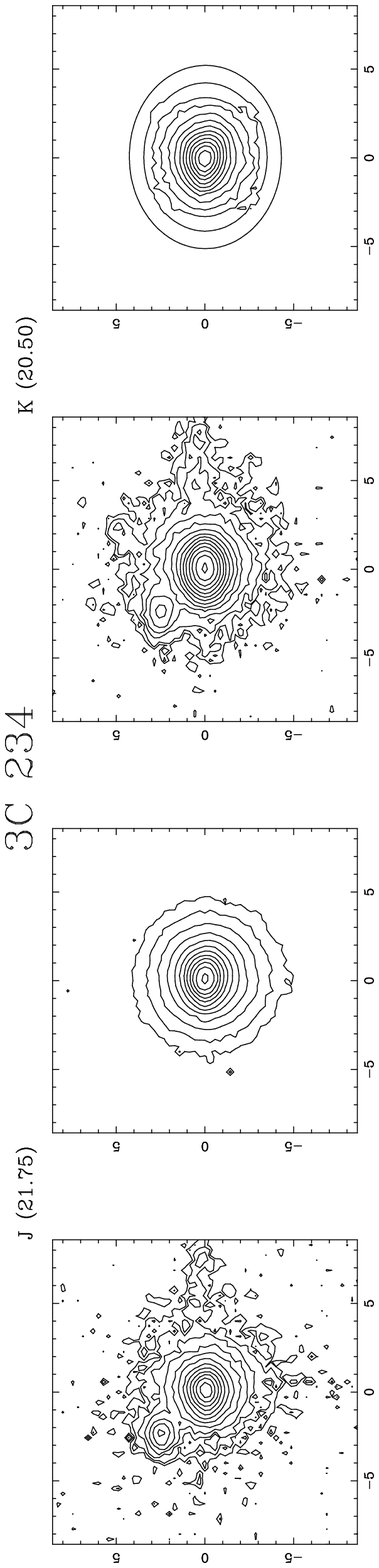}
\vspace*{48mm}
\caption[]{As Fig.~\ref{fig:cont33}, but for 3C~234.}
\label{fig:cont234}
\end{figure*}

For galaxies with large nuclear-to-host ratios (e.g.\ 3C~234) or large
angular effective radii (e.g.\ 3C~84), the magnitude of the nuclear
source was well-determined. However, for other sources it was quite
sensitive to the PSF used, and in many cases, the best-fit nuclear
flux was very faint and it became necessary to estimate some limit,
fainter than which a source would not be considered a true detection.
We estimate the uncertainty on the nuclear flux (in units of detector
counts) to be $\sigma = \overline{e} \sqrt{n}$, where $\overline{e}$
is the mean value of the error image in a 3-arcsec aperture and $n$ is
the number of pixels in this aperture. When the flux of the nuclear
source is less than $5\sigma$, we quote an upper limit. These limits
vary slightly from source to source, depending on such things as the
seeing, but are typically $J \approx 18.5$ and $K \approx 17.5$, or a
few per cent of the integrated host galaxy luminosity. The size of the
variations in the best fit value of the nuclear flux for different
PSFs and starting simplexes was broadly consistent with this level of
uncertainty. Contour plots of the data and our models can be found in
Figs~\ref{fig:cont33}--\ref{fig:cont234}. The properties of the
best-fitting models are listed in Table~\ref{tab:fitres}.

\begin{table}
\caption[fitres]{Results of the fits to the $J$ and $K$ images. Listed
are the nuclear and host galaxy (integrated out to infinite radius)
magnitudes, their ratio in a 12-arcsec aperture, and the effective
radius (in kpc), axial ratio and position angle of the host galaxy.}
\label{tab:fitres}
\centering
\begin{tabular}{l@{ }crrrrcrc}
Galaxy & & \cntr{1}{$m_{\rm nuc}$} & $m_{\rm host}$ &
$\frac{L_{\rm nuc}}{L_{\rm host}}\,(12'')$ & \cntr{1}{$r_{\rm e}$} &
$a/b$ & \cntr{1}{$\Theta$} & $\chi^2_\nu$ \\
\hline
3C~33  & $J$ & $>18.2$& 12.6 & $<0.01$ &  7.5 & 1.10 & 153 & 1.78 \\
       & $K$ &   16.6 & 11.7 &   0.02  &  5.9 & 1.09 & 158 & 2.05 \\
3C~42  & $J$ & $>19.0$& 16.4 & $<0.13$ & 15.1 & 2.32 & 160 & 1.58 \\
       & $K$ & $>18.1$& 14.9 & $<0.07$ & 12.9 & 1.95 & 162 & 1.01 \\
3C~79  & $J$ &   17.4 & 15.5 &   0.23  & 10.4 & 1.15 &  15 & 1.90 \\
       & $K$ &   15.6 & 14.2 &   0.35  &  8.0 & 1.07 &  11 & 1.27 \\
3C~84  & $J$ &   15.7 &  8.9 &   0.02  & 24.6 & 1.24 & 112 & 5.53 \\
       & $K$ &   12.6 &  7.9 &   0.15  & 25.2 & 1.26 & 114 & 8.25 \\
3C~98  & $J$ & $>18.5$& 11.9 & $<0.01$ &  4.2 & 1.17 &  60 & 2.72 \\
       & $K$ & $>17.2$& 10.8 & $<0.01$ &  4.5 & 1.16 &  62 & 1.27 \\
3C~153 & $J$ & $>18.4$& 15.4 & $<0.10$ & 21.5 & 1.45 &  86 & 0.14 \\
       & $K$ & $>17.5$& 14.0 & $<0.07$ & 21.5 & 1.37 &  87 & 0.22 \\
3C~171 & $J$ & $>18.8$& 15.9 & $<0.08$ &  7.0 & 1.21 & 170 & 1.24 \\
       & $K$ & $>17.8$& 14.5 & $<0.06$ &  8.6 & 1.33 & 156 & 1.88 \\
3C~192 & $J$ & $>18.4$& 13.0 & $<0.01$ &  7.8 & 1.03 &  94 & 1.83 \\
       & $K$ & $>17.4$& 12.1 & $<0.01$ &  7.7 & 1.04 & 101 & 1.65 \\
3C~223 & $J$ &   17.5 & 14.5 &   0.10  & 11.1 & 1.33 &  57 & 0.68 \\
       & $K$ &   14.9 & 13.6 &   0.50  & 13.7 & 1.42 &  59 & 1.32 \\
3C~234 & $J$ &   15.5 & 15.4 &   1.30  & 13.7 & 1.04 &  61 & 1.87 \\
       & $K$ &   13.6 & 13.8 &   1.66  & 12.2 & 1.21 &  81 & 1.17 \\
\hline
\end{tabular}
\end{table}

We make a comment regarding the $K$-band fit for 3C~98 listed in
Table~\ref{tab:fitres}. This fit has $\chi^2/\nu = 25226/19876$, but
the other nine fits all have $\chi^2 < 26000$ yet have nuclear sources
as bright as $K = 14.9$. This is in marked contrast to the other model
galaxies, where large variations in the best-fitting nuclear flux
compared to the value in Table~\ref{tab:fitres} were accompanied by
large increases in $\chi^2$, and were presumably due to a
poorly-matched PSF. We shall return to this point in
Section~\ref{sec:nonuke}.

\subsection{Thermal images}
\label{sec:lm}

We were able to detect the radio galaxies at $L'$ in eight out of ten
cases. We were less successful at $M$, and only detected three sources
of the eight which we observed. For most of our detections, the flux
is dominated by the transmitted quasar nucleus, although the stellar
contribution is not negligible.  The fact that we measure larger
fluxes in the larger apertures shows that we have detected starlight
from the host galaxies, but the signal-to-noise ratio per pixel is far
too low to allow the two-dimensional modelling used for the shorter
wavelength images.  Indeed, for most galaxies the extended emission is
not bright enough to even permit a meaningful measurement of the flux
in a circumnuclear annulus. We must therefore estimate the stellar
flux by extrapolating from our shorter wavelength images.

The stellar populations of early-type galaxies are dominated in the
near-infrared by stars on the red giant branch. We therefore use the
observed spectrum of the M2II--III star $\beta$~Peg (Strecker,
Erickson \& Witteborn 1979) as a model for the stellar populations of
our galaxies. The colours of $\beta$~Peg are known to match those of
early-type galaxies, and the star's spectrum has been used to compute
$K$-corrections for elliptical galaxies (see Longmore \& Sharples
1982). We correct the observed 3-arcsec $K$ magnitudes for the
presence of the nuclear source listed in Table~\ref{tab:fitres} to
determine the flux from stars only in this aperture, and use the
$\beta$~Peg spectrum to estimate the stellar contribution to the flux
at $L'$ and $M$. If there is an excess at greater than $3\sigma$
significance in the 3-arcsec aperture, we attribute it to the quasar
nucleus. The results of this analysis are presented in
Table~\ref{tab:nukemag}.

\begin{table}
\caption[]{Inferred magnitudes for the nuclear source in each of the radio
galaxies.}
\label{tab:nukemag}
\centering
\begin{tabular}{lrrrr}
Galaxy & \cntr{1}{$J$} & \cntr{1}{$K$} & \cntr{1}{$L'$} & \cntr{1}{$M$} \\
\hline
3C~33  & $>18.20$ &   16.57  &   12.32  &   11.33 \\
3C~42  & $>19.00$ & $>18.10$ & $>15.82$ & \cntr{1}{$\ldots$} \\
3C~79  &   17.37  &   15.89  &   13.11  & $>12.54$ \\
3C~84  &   15.70  &   12.59  &    9.73  & \cntr{1}{$\ldots$} \\
3C~98  & $>18.50$ & $>17.20$ & $>13.10$ & $>13.46$ \\
3C~153 & $>18.40$ & $>17.50$ & $>16.58$ & $>12.86$ \\
3C~171 & $>18.80$ & $>17.80$ & $>16.58$ & $>12.51$ \\
3C~192 & $>18.40$ & $>17.40$ & $>14.99$ & $>12.63$ \\
3C~223 &   17.52  &   14.91  &   12.98  &   11.81 \\
3C~234 &   15.54  &   13.65  &   10.56  &    9.94 \\
\hline
\end{tabular}
\end{table}

Although 3C~98 and 3C~192 are both detected at $L'$, the emission we
observe is purely stellar in origin, as both sources have 3-arcsec
aperture $K-L'$ colours consistent with the $\beta$~Peg spectrum,
although the colour of 3C~98 is redder than $\beta$~Peg's and only
marginally consistent with it. However, Lilly et al.\ (1985a) showed
that even in a much larger aperture (7.5\,arcsec), 3C~98 is redder
than a typical early-type galaxy, which rules out a compact red source
as an explanation for the colour. We note that the $L'-M<-0.17$ limit
is also consistent with starlight, but is far too blue for a quasar
(corresponding to a power law index $\alpha < -2.5$).

\subsection{Reality of our detections}

\begin{figure}
\includegraphics{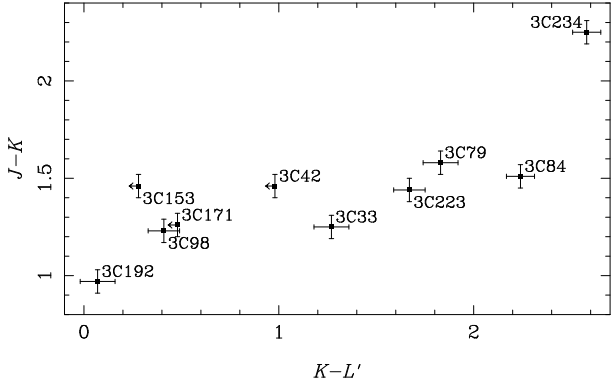}
\vspace*{54mm}
\caption[]{Two colour diagram for the ten radio galaxies. Colours have
been measured in a 3-arcsec aperture.}
\label{fig:colcol}
\end{figure}

If we examine Fig.~\ref{fig:colcol}, which presents the 3-arcsec
aperture colours of the ten radio galaxies from Table~\ref{tab:phot},
the galaxies split naturally into two groups: those with $K-L'$
colours consistent with late-type stellar populations, and those which
are much redder. There is also a weak, yet highly significant,
correlation between the two colours, as expected from a red nuclear
component which becomes increasingly significant at $K$. The simple
colour analysis of Section~\ref{sec:lm} will obviously infer a nuclear
source for galaxies in the second group only, but it is important to
note that these are also the only galaxies for which nuclear sources
were detected in Section~\ref{sec:jk}. All nuclear sources which are
reliably detected at $K$ in our 2D analysis should be detectable in
our $L'$ images due to their red colour; this is indeed true.
Similarly, any real nuclear sources we detect at $J$ should also be
seen at $K$ (and therefore at $L'$ too), and this is also the case. A
similar argument does not apply to our $M$-band data, since the
reduction in sensitivity over $L'$ ($\sim 3$\,mag) is much greater
than can be compensated for by the red colour of the quasar nucleus
($L'-M=0.76$ for a power law $\alpha=1.3$; Neugebauer et al.\
1987). We are therefore convinced that we have not made any spurious
detection, and that we have not failed to detect any sources which are
in reality brighter than our quoted limits.

\section{Luminosity and extinction estimates}

\subsection{Galaxies with detected nuclei}
\label{sec:extinction}

\begin{figure}
\includegraphics{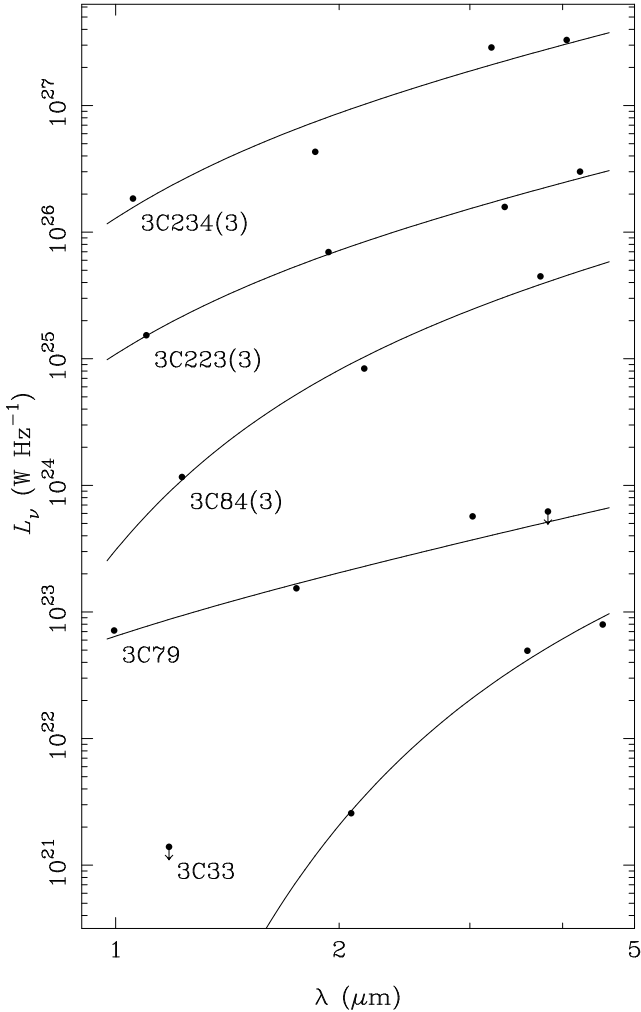}
\vspace*{135mm}
\caption[]{Rest-frame nuclear luminosities and best-fit reddened quasar
models for the five radio galaxies in which a nuclear source was detected.
For clarity some of the galaxies have been shifted upwards in the
$y$-direction by the number of dex given in parentheses.}
\label{fig:irspec}
\end{figure}

If the standard interpretation of our nuclear sources as dust-obscured
quasars is correct, they should have the colours of a quasar, modified
by some amount of foreground reddening. We therefore perform linear
regression, accounting for upper limits in the manner of Isobe,
Feigelson \& Nelson (1986), on the data of Section~\ref{sec:nuclei}.
This analysis produces values for the intrinsic luminosity and
extinction of each nuclear source. We adopt our own parametrization of
the near-infrared interstellar extinction law, obtained by fitting a
second-order polynomial to the data of Rieke \& Lebofsky (1985). This
parametrization,
\[
A_\lambda/A_V = 0.243\lambda^{-2} + 0.184\lambda^{-1} - 0.022 ,
\]
by virtue of ignoring the optical data, provides a rather better fit
to the longest wavelengths than do those of Howarth (1983) and
Cardelli, Clayton \& Mathis (1989). The intrinsic quasar spectrum is
assumed to be that of an $\alpha = 1.3$ power law (Neugebauer et al.\
1987). Table~\ref{tab:exts} lists the extinctions and intrinsic
luminosities we derive for the obscured quasars in the five radio
galaxies with detected nuclear sources.  These data are presented
graphically in Fig.~\ref{fig:irspec}. For 3C~223 and 3C~234, where the
nucleus was detected at all four wavelengths, we also performed
regression with the spectral index as an additional free parameter. In
both cases the spectral index was found to be close to 1.3 and the
results were not changed significantly.

\begin{table}
\caption[]{Foreground extinction and unobscured rest-frame 1\,$\mu$m
luminosity for the obscured quasar nuclei detected in our analysis. The
dereddened \OIII~$\lambda$5007 emission line luminosity from
Table~\ref{tab:sample} is also listed.}
\label{tab:exts}
\centering
\begin{tabular}{lr@{.}lr@{.}lc}
Galaxy & \cntr{2}{$A_V$ (mag)} & \cntr{2}{$\log L_{\rm1\,\mu m}$
(W\,Hz$^{-1}$)} & \cntr{1}{$\log L_{5007}$ (W)} \\
\hline
3C~33  & 29&$5 \pm 3.5$ & \hspace*{5mm} 22&$49 \pm 0.11$ & 35.23 \\
3C~79  &  1&$0 \pm 2.5$ & \hspace*{5mm} 23&$07 \pm 0.25$ & 36.19 \\
3C~84  &  9&$4 \pm 1.6$ & \hspace*{5mm} 22&$02 \pm 0.12$ & 35.07 \\
3C~223 &  3&$9 \pm 0.6$ & \hspace*{5mm} 22&$72 \pm 0.05$ & 35.53 \\
3C~234 &  4&$0 \pm 1.9$ & \hspace*{5mm} 23&$83 \pm 0.15$ & 36.47 \\
\hline
\end{tabular}
\end{table}

\subsection{Galaxies without detected nuclei}
\label{sec:nonuke}

The five galaxies with detected nuclei will obviously preferentially
include those objects with low nuclear extinction. If we wish to reach
some conclusions about the amount of obscuration in a `typical' radio
galaxy, we must make estimates of the extinction in the remaining five
members of the sample. This requires us to calculate the intrinsic
near-infrared luminosity of the quasars in these objects, which we do
in the following manner.

We first estimate the unobscured optical luminosity of the quasar
using the \OIII~$\lambda$5007 emission line. Simpson (1998a) has
argued that this line is a good indicator of the intrinsic luminosity
of radio galaxies and lobe-dominated quasars (such as those selected
at low radio frequency). It has also been shown that the \OIII\
luminosities of lobe- and core-dominated quasars are the same for a
given extended radio luminosity (Jackson et al.\ 1989; Jackson \&
Browne 1991; Corbin 1997), indicating that the line is emitted
isotropically.

There is no Baldwin effect in the equivalent width of \OIII\ in
lobe-dominated quasars (Jackson \& Browne 1991). Only when
core-dominated objects are considered does such an effect appear. This
can be understood as the result of an increase in the continuum
luminosity at small polar angles due to an anisotropic component.
However, since the near-infrared emission is believed to be produced
by dust heated by the optical--ultraviolet radiation and located close
to the equatorial plane of the quasar, it is the luminosity at large
viewing angles, such as is seen in lobe-dominated quasars, which is of
interest to us.

By virtue of the above arguments, we can use the optical--infrared
properties of lobe-dominated quasars as benchmarks for the properties
of the quasars within our radio galaxies. The mean \OIII\ equivalent
width for the $R<1$ quasars from Jackson \& Browne (1991) is $\log
{\rm EW}_{5007}\,({\rm\AA}) = 1.66 \pm 0.32$, or 46\,\AA\ with a
factor of 2 scatter. This is larger that the equivalent width of
24\,\AA\ determined by Miller et al.\ (1992) for low redshift
($z<0.5$) quasars selected from the Palomar--Green Bright Quasar
Sample (BQS; Schmidt \& Green 1983), because optically selected
samples like the BQS preferentially contain objects oriented close to
the line of sight where the optical continuum is boosted. This is
supported by the large values of the 5-GHz core-to-lobe ratio, $R$,
displayed by the radio-loud quasars in the low redshift BQS (Miller,
Rawlings \& Saunders 1993).

Adopting 46\,\AA\ as the rest-frame equivalent width (with respect to
the non-stellar continuum) of the \OIII~$\lambda$5007 emission line,
and assuming $\alpha = 0.4 \pm 0.4$ as representative of the continuum
between 5007\,\AA\ and 1\,$\mu$m (Neugebauer et al.\ 1987), we find
that the last two columns of Table~\ref{tab:exts} should be related by
\[
\log L_{\rm1\,\mu m} - \log L_{5007} = 12.62 \pm 0.34 \, ,
\]
which is indeed the case, and indicates the validity of our method.

Applying this technique to the five radio galaxies without detected
nuclear sources, we derive lower limits to the nuclear extinction. The
non-detections of compact sources in 3C~42 and 3C~153 do not rule out
unobscured nuclei with any significance, since both these sources have
faint emission line fluxes and therefore are expected to have quite
faint nuclei. For the other three galaxies we determine the following
$3\sigma$ limits: $A_V > 74^m$ for 3C~98; $A_V > 37^m$ for 3C~171; and
$A_V > 45^m$ for 3C~192. Note that these limits are derived from the
non-detections at $M$-band (where our sensitivity to highly obscured
nuclei is greatest) and are therefore independent of the $J$ and
$K$-band fits. The high extinction towards the nucleus of 3C~98
therefore clearly precludes a $K$-band nucleus brighter than the limit
we quote in Table~\ref{tab:fitres}.

\section{The geometry of the obscuring material}

\begin{figure}
\includegraphics{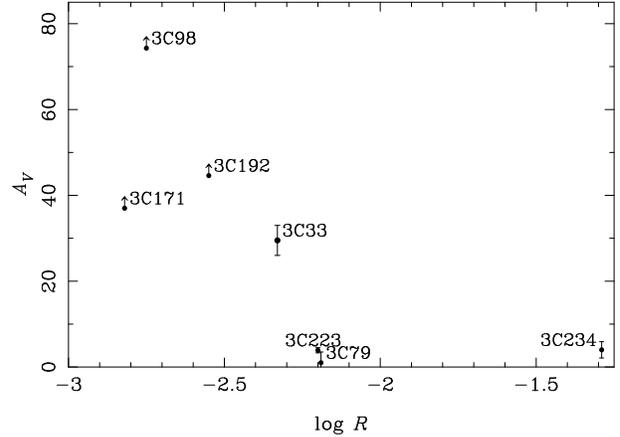}
\vspace*{60mm}
\caption[]{Inferred nuclear extinction for the galaxies in our sample,
plotted against the core-to-lobe ratio, $R$. An anti-correlation is
readily apparent.}
\label{fig:correl}
\end{figure}

The core dominance parameter, $R$ (also called the core-to-lobe
ratio), is a widely-used orientation indicator (Orr \& Browne 1982)
and we are therefore capable of determining the nuclear extinction to
our radio galaxies as a function of viewing angle. Simpson (1994a,
1996) proposed analysing a large sample of radio galaxies in this way
to distinguish between different geometries for the obscuring
material, which partly provided the motivation for the study described
here. The value of $R$ is usually determined at 5\,GHz, and the
properties of our sources at this frequency are listed in
Table~\ref{tab:cores} of the Appendix. We omit 3C~84 since it does not
possess the characteristic FR\,II radio morphology. The values of the
nuclear extinction derived earlier are plotted against $\log R$ in
Fig.~\ref{fig:correl}.

Using a generalized Kendall's rank correlation statistic (Isobe et
al.\ 1986), we can reject the hypothesis that $A_V$ is uncorrelated
with $R$ at better than 95\% confidence, even if 3C~234 is excluded
from the analysis. The nuclear extinction therefore increases with
viewing angle (i.e.\ as $R$ decreases) and suggests a flattened
distribution for the obscuring material.

Ignoring density inhomogeneities, the obscuring material's geometry is
defined by $A_V(\theta)$, which we can derive from $A_V(R)$ if we can
transform the observed core dominance parameter, $R$, to a viewing
angle, $\theta$. These two quantities are related by
\[
R = \frac{1}{2}R_{\rm T} (1 - \beta \cos\theta)^{-2} + \frac{1}{2} R_{\rm
T} (1 + \beta \cos\theta)^{-2}
\]
(Orr \& Browne 1982), where $R_{\rm T} \equiv R(90\degr)$ and $\beta$
is the velocity of the beamed core component relative to the speed of
light. We adopt $\beta = 0.99$ (Orr \& Browne 1982), although its
exact value is unimportant except at small values of $\theta$, which
are not relevant to our study of radio galaxies.

To convert from $R$ to $\theta$, it is necessary to know the value of
$R_{\rm T}$. Orr \& Browne determined $R_{\rm T} = 0.024$ from a
sample of 3CR quasars, but they assumed that these objects were
randomly oriented with respect to the line of sight. Since it is now
believed that quasars are preferentially pointed towards us, Orr \&
Browne's value of $R_{\rm T}$ will be an overestimate. We therefore
redetermine $R_{\rm T}$ using both quasars and FR\,II radio galaxies
from Laing et al.\ (1983) with $z<0.43$, although we exclude 3C~236
since it possesses a steep-spectrum radio core. As described in the
Appendix, a good fit to the data is obtained with $P(\log R_{\rm T}) =
{\sf N}(-2.54,0.51)$. This is consistent with the distribution derived
by Simpson from the radio galaxies alone, and with estimates made from
other samples of radio galaxies (e.g.\ Morganti et al.\ 1997).

\begin{figure}
\includegraphics{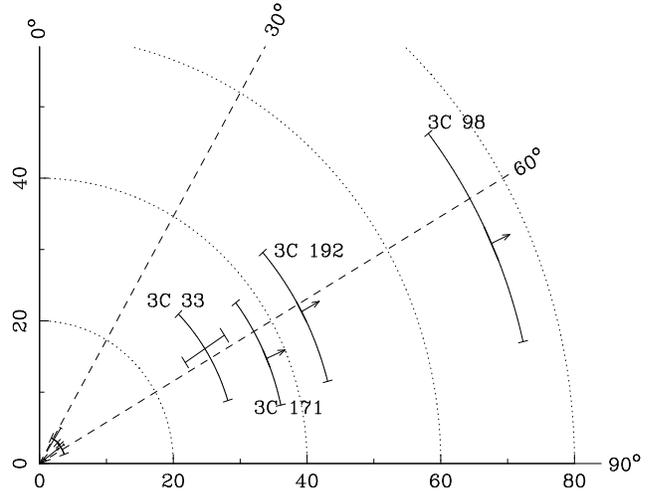}
\vspace*{64mm}
\caption[]{A cross section through the torus perpendicular to the
equatorial plane. The axes are in units of magnitudes of visual
extinction. The dotted lines describe the polar coordinate frame and
are there to guide the eye.}
\label{fig:torus}
\end{figure}

Since we have only determined the {\em a priori\/} probability
distribution for $R_{\rm T}$, and not its actual value for each radio
galaxy, the observed core dominance parameter, $R$, is transformed
into a probability distribution for the viewing angle, $\theta$. This
transformation is described in the Appendix. We list the viewing
angles for the radio galaxies in Table~\ref{tab:theta} and plot the
cross-section through the torus in Fig.~\ref{fig:torus}. It is
immediately clear from this analysis that our inability to constrain
the angle at which we are viewing each object is a major limitation to
the effectiveness of this technique. Doppler boosting only produces a
fourfold increase in the observed luminosity of the central component
for $\theta \approx 50\degr$ (the viewing angle which is believed to
separate quasars from radio galaxies at low redshift; L94, Simpson
1998a, and see also Table~\ref{tab:theta}), barely enough to produce a
signal over the factor of three scatter in $R_{\rm T}$. Since core
dominance is believed to be the best available orientation indicator,
it is clear that a much larger sample of radio galaxies will be needed
to produce a useful torus map.

\begin{table}
\caption[]{Core dominance parameters and viewing angles with respect
to the radio axis for the radio galaxies in our sample.}
\label{tab:theta}
\begin{center}
\begin{tabular}{lrr@{}l}
Galaxy & \cntr{1}{$\log R$} & \cntr{2}{$\theta$ (\degr)} \\
\hline
3C~33  &  $-2.33$ &   57 &$^{+16}_{-12}$ \\
3C~42  &  $-2.69$ &   64 &$^{+11}_{-13}$ \\
3C~79  &  $-2.19$ &   53 &$^{+18}_{-11}$ \\
3C~98  &  $-2.75$ &   65 &$^{+12}_{-14}$ \\
3C~153 & $<-3.52$ & $>54$& ~($2\sigma$)  \\
3C~171 &  $-2.82$ &   67 &$^{+10}_{-15}$ \\
3C~192 &  $-2.55$ &   62 &$^{+13}_{-14}$ \\
3C~223 &  $-2.20$ &   54 &$^{+17}_{-12}$ \\
3C~234 &  $-1.29$ &   33 &$^{+16}_{-7}$  \\
\hline
\end{tabular}
\end{center}
\end{table}

\section{The $K$--$z$ relation at low redshift}

\begin{figure}
\includegraphics{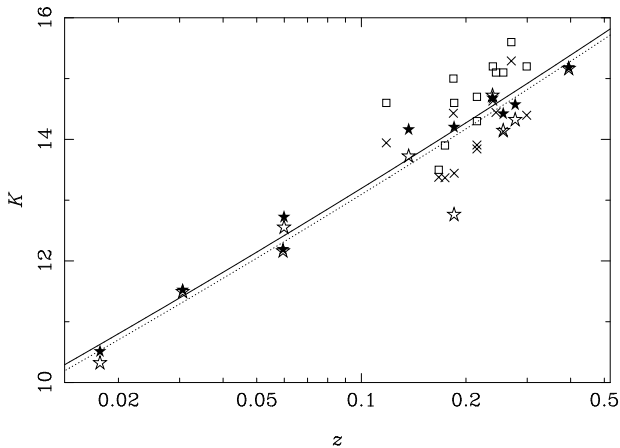}
\vspace*{64mm}
\caption[]{The $K$-band Hubble diagram for low redshift radio galaxies.
Open stars are the observed magnitudes for our 3CRR radio galaxies,
filled stars are the model host galaxy magnitudes for these objects,
and crosses and open squares are the observed and model host galaxy
magnitudes for T96's radio galaxies (all measured or synthesized in
12-arcsec apertures). The dotted line is the $K$--$z$ relation from
Lilly et al.\ (1985b), and the solid line is this relation shifted by
0.10\,mag to provide the best least-squares fit to our model
magnitudes.}
\label{fig:kz}
\end{figure}

T96 claimed that the $K$ magnitudes of their radio galaxy hosts are
0.5\,magnitudes fainter than the currently-assumed $K$--$z$ relation,
with the discrepancy being caused by the contribution from the quasar
nucleus. They detected nuclear sources in all twelve of their
galaxies, with a mean nuclear-to-host ratio $\langle L_{\rm nuc} /
L_{\rm host} \rangle = 0.80$. Our study, on the other hand, failed to
detect a quasar contribution in five of our objects, despite a
similarly faint detection threshold. What effect does our result have
on the $K$--$z$ locus? We find the best least-squares fit to our data
that can be obtained by applying an offset, $\Delta$, to the $K$--$z$
relation of Lilly et al.\ (1985b),
\[
K = 49.87 - \sqrt{1092.7 - 259.74 \log z} + \Delta .
\]
For the raw data, the best offset is effectively zero ($\Delta =
-0.01$), as expected given the accuracy of our photometry and the
unbiased nature of our sample. If we consider instead just the host
magnitudes of Table~\ref{tab:fitres}, we find $\Delta = 0.10$,
supporting a shift of the $K$--$z$ relation to fainter magnitudes, but
by a much smaller amount than that proposed by T96. In addition, this
should probably be considered an upper limit to any shift required in
the $K$--$z$ relation, since sources from fainter radio surveys (which
lie on same $K$--$z$ relation due to the lack of a correlation between
radio and host galaxy luminosity and low redshift) will have
intrinsically fainter infrared nuclei, which should therefore
contribute less flux. We display our results in Fig.~\ref{fig:kz},
together with the host galaxy magnitudes derived by T96, which are
clearly fainter on average. We investigate this discrepancy, first by
addressing the difference in the detection rates for nuclear sources
between our study and that of T96, and then by looking at the
difference in host galaxy magnitudes. We assume that the nuclear
sources we and T96 have detected are in fact active nuclei, based on
their luminosities and broad-band SEDs (T96 make a similar claim). If
instead they are merely stellar cusps (albeit ones with unusual
colours), then there is obviously no reason to shift the $K$--$z$
relation at all, since the actual form of the radial surface
brightness profile is unimportant.

\subsection{Detection efficiency}

While our sample is effectively complete (only the random exclusion of
3C~123 prevents it from being truly so), the twelve objects studied by
T96 were selected to match previously-constructed samples of
radio-loud and radio-quiet quasars. One criterion was a match in radio
spectral index, and T96's sample therefore included a number of radio
galaxies with spectra much flatter than the typical $\alpha \approx
0.7$. There are five such flat-spectrum radio galaxies in the sample
of T96, and they are therefore strongly overrepresented compared to
the general low frequency selected radio galaxy population (there are
none in our sample). Flat integrated radio spectra are produced when
the flat spectrum radio core dominates over the steep spectrum lobes
in the frequency region of interest, and consequently such sources
have high values of the core-to-lobe parameter, $R$.
Fig.~\ref{fig:correl} showed the strong anti-correlation between $A_V$
and $R$, which we interpreted in terms of the geometry of the
obscuring material. We should therefore expect the nuclei of T96's
radio galaxies to be less heavily obscured than those of our sample
members. This should result in brighter (and hence more readily
detectable) nuclei and a tendency for their sources to lie below
(i.e.\ be brighter than) the $K$--$z$ relation. Indeed, an offset of
0.14\,mag to brighter magnitudes needs to be applied to the relation
of Lilly et al.\ (1995b) in order to obtain the best fit to their raw
data.

The single-filter imaging of T96 does not permit the detailed method
of extinction determination we employed in
Section~\ref{sec:extinction}.  Instead we estimate the average
extinction for T96's radio galaxy sample as a whole, by comparing the
luminosities of the nuclei detected in the radio galaxies with those
detected in the radio-loud quasars. Due to the way the samples were
matched, the intrinsic luminosities of the active nuclei should be the
same. Their Tables~4 and 5 indicate that the mean observed luminosity
of the radio galaxy nuclei is 1.77\,mag fainter than that of the
radio-loud quasar nuclei. A similar result is found when the $L_{\rm
nuc}/L_{\rm host}$ ratios are compared. For a typical redshift $z
\approx 0.2$, this corresponds to $A_V \approx 13^m$. By contrast, our
sample has an average extinction $A_V > 23^m$.

Our estimate appears to be more in line with the admittedly very
limited previous study in this field. Simpson (1994a) studied a small,
but complete, sample of nearby radio galaxies from the MS4 survey
(Burgess \& Hunstead 1994) which included three Class~A (Hine \&
Longair 1979) objects. A nuclear source was detected in one object,
which further study revealed to suffer an extinction $A_V = 30 \pm
4^m$ (PKS~0634$-$205; Simpson et al.\ 1995). The lower limits
determined for the other two Class~A sources implied an average $A_V >
21^m$. The powerful radio galaxy Cygnus~A has also had its quasar
nucleus detected in the near-infrared (Djorgovski et al.\ 1991), and
the most recent extinction estimates indicate very heavy obscuration
($A_V \approx 150^m$; Simpson 1994a, Ward 1996).

We also note that two of T96's radio galaxies, 3C~219 (L94; Hill et
al.\ 1996) and 3C~287.1 (Eracleous \& Halpern 1994) clearly show broad
H$\alpha$ in their optical spectra. It is therefore to be expected
that these two objects possess bright nuclear sources in the
near-infrared. We conclude that T96's sample is predisposed towards
sources with relatively low nuclear obscuration, and believe that our
sample, being effectively complete and free from selection biases, is
more representative of the radio galaxy population at large.

\subsection{Host galaxy magnitudes}

Even though T96's sample of radio galaxies appears to be biased
towards objects with low nuclear obscuration, the host galaxies should
not be affected. Our multi-wavelength images allow us to compute
robust upper limits to the non-stellar flux at $K$ (and hence an upper
limit to the shift of the $K$--$z$ relation required) in a relatively
simple manner which does not involve a ``black box'' routine such as
the two-dimensional fitting procedure. First, we assume that all the
$L'$ flux in a 3-arcsec aperture arises from the nucleus -- since
there {\em must\/} be some starlight present, this will overestimate
the true nuclear flux. In the three instances where no detection was
made at $L'$, we use the $3\sigma$ upper limit (again, overestimating
the true flux). We then determine an upper limit to the $K$-band flux
by assuming that the spectrum is an unreddened $\alpha = 1.3$ power
law. Although there will be variations in the near-infrared spectral
indices of the objects in our sample, these should average out among
the ten galaxies. We compute the faintest possible host galaxy
$K$-band magnitude by subtracting this nuclear flux from the observed
12-arcsec aperture flux. When this analysis is applied to 3C~234, the
maximum nuclear strength we derive exceeds the observed 12-arcsec
magnitude, which indicates the conservative nature of our
approach. Obviously, we cannot merely exclude 3C~234 from the
analysis, since it is the most strongly contaminated source in our
sample, so we assume the host has $K = 14.2$, as determined by our
model-fitting (T96 obtain a very similar result). Even with these
gross overestimates of the nuclear fluxes (and hence underestimates of
the host galaxy brightnesses), the offset from the $K$--$z$ relation
is only $\Delta = 0.35$, somewhat less than the 0.5\,mag found by
T96. More realistic estimates, such as assuming that the nucleus
suffers at least 5\,mag of extinction (just enough to obscure broad
H$\alpha$), imply $\Delta < 0.27$. The $K$--$z$ relation followed by
the host galaxies in our sample cannot therefore be the same as that
found by T96 for the galaxies in their sample, unless the nuclei of
our radio galaxies have intrinsic near-infrared spectral indices of
$\alpha < 0.2$. This is obviously incompatible with unification models
which claim that the nuclei of radio galaxies are identical to those
of quasars.

How then did T96 find such a large shift in the $K$--$z$ relation? We
first rule out the possibility that the galaxies in T96's sample are
genuinely fainter than ours. Although T96's sample is generally of
lower radio luminosity than ours, Hill \& Lilly (1991) have shown that
there is no difference in the optical magnitudes of radio galaxies
over three orders of magnitude in radio luminosity, and the
optical--infrared colours of the two samples cannot differ by half a
magnitude. In addition, Eales et al.\ (1997) have shown that the $K$
magnitudes of radio galaxies from the 3C and 6C samples (6C is
selected at a $\sim 5$ times fainter flux level) are similar at low
redshift. Finally, if we exclude those radio galaxies from T96's
sample which do not meet the 3CRR flux limit, their result is
virtually unchanged. Similarly, if we consider only FR\,II galaxies
from T96, or only those galaxies with a steep radio spectrum, we find
no significant change to their result.

We consider two likely explanations for the different $K$--$z$
relations inferred. We first consider what effect the small field of
view ($\sim 35\arcsec$) of IRCAM1 might have had on T96's modelling.
While the much larger field of view of IRCAM3 provides a significant
amount of blank sky in our images and therefore allows an accurate
determination of the background level, this was not the case with
IRCAM1. As T96 explain, they may have overestimated the sky level in
their images, and we discuss here why this can lead to an
underestimation of the host galaxy brightness. T96's radio galaxies
have a typical effective radius of a few arcseconds. At the edge of
their field of view, a de Vaucouleurs profile does not have a
negligible surface brightness, but rather will have $\mu \approx
23$\,mag\,arcsec$^{-2}$ (for a typical $\mu_{1/2} \approx 20.5$ and
$r_{\rm e} \approx 6\arcsec$). We have investigated the effects of
this on modified $K$-band images of some of our galaxies, by
subtracting a constant value from all the pixels to simulate
overestimation of the sky level.

We find that for small deviations from the true value, the effective
radius of the best-fitting model does not change significantly, but
the flux normalization of the stellar component is decreased. This
normalization is primarily determined at $r \approx r_{\rm e}$, since
the signal-to-noise ratio is much lower at larger radii, while the
sampling errors can be large near the centre, and the flux of the
nuclear source can be varied to compensate for deficiencies in the
quality of the fit. For a typical galaxy with an error in background
determination of the level indicated above, we find an increase of
$\sim 20$--40\% increase in the flux attributed to the nucleus. The
value of the $\chi^2$ statistic also increases, and it is therefore
worth considering whether more accurate results might be obtained if
the sky level is included as an extra fitting parameter in the
modelling process.

There is also evidence that T96 may have systematically overestimated
the scale lengths of their galaxies. Optical imaging with the {\it
Hubble Space Telescope\/} (Dunlop et al.\ 2000) produced a median
scale length of 12\,kpc, compared to 20\,kpc determined by T96. The
effect of overestimating the $r_{\rm e}$ is also to lower the
contribution from the host galaxy in the central regions, and
therefore to overestimate the contribution from a nuclear point
source. A systematic error of $\sim +70$\% in the scale lengths
determined by T96 would serve to produce central surface brightnesses
$\sim 1$\,mag too faint (again, the fitting procedure will tend to
match the model to the data at $r \approx r_{\rm e}$) and the model
flux within a 12\arcsec\ aperture will be underestimated by $\sim
0.4$--0.7\,mag, depending on the size of the galaxy. Since this is
very much in line with the shift determined by T96, it would seem to
be a likely explanation for their result.

We can make a direct comparison between our results and those of T96
for the two objects which are common to both samples. We present the
results of our $K$-band fits and those of T96 in
Table~\ref{tab:compare}. The agreement for 3C~234 is excellent, but it
is less good for 3C~79. Optical imaging with the {\it Hubble Space
Telescope\/} (Dunlop et al.\ 2000) produces an effective radius
$r_{\rm e} = 9.4$\,kpc, somewhere between the values we determine from
our $J$ and $K$-band images, and clearly indicating that T96
overestimated the true value. A visual inspection of their Fig.~A2
appears to indicate that the larger effective radius is influenced by
the companion source approximately 6\arcsec\ north of the radio
galaxy. Alternatively, there may have been additional scattered light
within IRCAM1 which would artificially enhance the surface brightness
at large radial distances and hence cause the scale length to be
incorrectly overdetermined, although the good agreement for 3C~234
(where the bright nuclear point source would be expected to produce a
large effect) seems to argue against this explanation.

\begin{table}
\caption[compare]{Direct comparison between our results and those of the
elliptical galaxy fits of Taylor et al.\ (1996). Host galaxy magnitudes
are within a simulated 12-arcsec aperture.}
\label{tab:compare}
\centering
\begin{tabular}{llcccrcr}
Object & & $K_{\rm nuc}$ & $K_{\rm host}$ & $\frac{L_{\rm nuc}}{L_{\rm
host}}$ & \cntr{1}{$r_{\rm e}$} & $a/b$ & \cntr{1}{$\Theta$} \\
\hline
3C~79  & Us  & 15.9 & 14.4 & 0.35 &  8.0 & 1.07 &  11 \\
       & T96 & 15.2 & 14.6 & 0.57 & 16.6 & 1.16 &   7 \\
3C~234 & Us  & 13.6 & 14.2 & 1.66 & 12.2 & 1.21 &  81 \\
       & T96 & 14.0 & 14.3 & 1.41 & 13.0 & 1.22 &  55 \\
\hline
\end{tabular}
\end{table}

It is quite possible that different effects are at work for different
objects. Since we can think of no likely effect which would serve to
{\em overestimate\/} the host galaxy brightnesses, we can expect a
significant systematic error in the results of T96, even if we cannot
be certain what the dominant cause of this error might be.

\section{Summary}

We have analysed near-infrared {\it JKL$'$M\/} images of an
effectively complete sample of ten 3CRR radio galaxies with strong
emission lines, and have separated them into their nuclear and stellar
components. We find that the colours of the nuclear components are
well-modelled by a reddened power-law, and the derived unobscured
luminosities are in excellent agreement with the values predicted from
the emission-line luminosities.

We find evidence that the nuclear extinction increases with viewing
angle, suggesting a flattened distribution for the obscuring material.
The contributions from the nuclear components make the model host
galaxy magnitudes fainter than the observed $K$ magnitudes, and
support a shift of the $K$--$z$ relation to fainter magnitudes by
0.1\,mag for low-redshift 3CRR galaxies. Using our observed $L'$
magnitudes and the assumption that the nuclear sources in radio
galaxies are intrinsically no bluer than quasar nuclei, we can rule
out a shift of more than 0.3\,mag, independent of our two-dimensional
model-fitting procedure.

We have investigated possible causes of the different results obtained
by Taylor et al.\ (1996) and ourselves. While we believe that T96's
sample is biased towards sources with low nuclear obscuration and
therefore a relatively large nuclear fraction, this should not affect
their host galaxy magnitudes. We believe that their fainter magnitudes
are caused by overestimating the sky level in their images and/or the
effective radii of their galaxies. We conclude that the magnitudes of
low-redshift radio galaxy hosts have been systematically overestimated
by no more than about 0.1\,mag as a result of non-stellar radiation
from the obscured quasar nuclei, and that conclusions drawn so far
concerning the cosmic evolution of radio galaxies (e.g.\ Lilly \&
Longair 1984; Eales \& Rawlings 1996) are therefore reliable.

\section*{Acknowledgments}

We are grateful to Jo McAllister for useful discussions regarding the
accuracy of the 2-D fitting results, and to the referee, Jim Dunlop,
for suggesting improvements to the manuscript. The United Kingdom
Infrared Telescope is operated by the Joint Astronomy Centre on behalf
of the U. K. Particle Physics and Astronomy Research Council, and we
thank the UKIRT staff for their help. We also thank Mike Goad for
providing a coded version of the simulated annealing minimization
routine. This work has made use of the NASA/IPAC Extragalactic
Database (NED), operated by the Jet Propulsion Laboratory, California
Institute of Technology, under a contract with the National
Aeronautics and Space Administration.

\appendix
\section{Core dominance and viewing angle}

This Appendix describes how we have used the observed core dominance
parameter of a radio galaxy to infer the viewing angle between its
radio axis and our line of sight. To do this, we have compiled in
Table~\ref{tab:cores} radio data for all FR\,II sources with $z<0.43$
in Laing et al.\ (1983). In the interests of keeping a homogeneous
dataset, we use the 5-GHz flux measurements and spectral indices
between 2.7 and 5\,GHz from Pauliny-Toth \& Kellerman (1968). Although
these are not necessarily the most accurate data available, this is
not a concern as our assumption of a common spectral index $\alpha =
0$ for the core components (which may be somewhat variable) will
introduce some errors in our determination of the rest-frame 5-GHz
core dominance parameter. For the four non-3C sources which were not
observed by Pauliny-Toth \& Kellerman, we have opted to use the data
from Herbig \& Readhead (1992). These authors perform fits to the
radio spectra of sources, which should also help to homogenize the
data they compile from the literature. Herbig \& Readhead list
luminosities and spectral indices at a rest-frame frequency of
2.5\,GHz, which we use to compute a rest-frame 5-GHz flux. However,
for consistency in Table~\ref{tab:cores}, we list synthetic {\em
observed-frame\/} 5-GHz fluxes determined from their fits. Note that
although we have included Hine \& Longair (1979) Class~B radio
galaxies in this sample to improve the statistics (the published
spectroscopy of the late-RA sources is too poor to make completely
accurate classifications and we would therefore have to exclude all of
them), their $R$ distribution is not significantly different from
that of the Class~A objects (a result also found by Morganti et al.\
1997).

\begin{table}
\caption[]{Properties of 3CRR FR\,II radio sources with $z < 0.43$. The
total 5-GHz fluxes and spectral indices between 2.7 and 5\,GHz are
from Pauliny-Toth \& Kellerman (1968) except where an asterisk ($^*$)
denotes synthesized data from the spectral fits of Herbig \& Readhead
(1992). We preferentially use 5-GHz core flux measurements, where
available, and assume a flat spectrum to convert measurements at other
frequencies. The last column identifies which paper in the reference
list was used to determine the core flux. Fluxes cited as from LP
(Leahy \& Perley 1991) are estimated from their radio maps.}
\label{tab:cores}
\begin{center}
\begin{tabular}{lrrrrrl}
\cntr{1}{Name} & \cntr{1}{$z$} & \cntr{1}{$S_{\rm5\,GHz}^{\rm tot}$}
& \cntr{1}{$\alpha^5_{2.7}$} & \cntr{1}{$S^{\rm core}$} &
\cntr{1}{$\log R$} & Core \\
& & \cntr{1}{(Jy)} & & \cntr{1}{(mJy)} & & ref \\
\hline
3C~16        & 0.4050 &  0.51 & 0.99 & $<0.5$ & $<-3.15$ & LP \\
3C~20        & 0.1740 &  4.18 & 0.70 &   2.6  &  $-3.25$ & FBP \\
3C~33        & 0.0595 &  5.03 & 0.57 &  24.0  &  $-2.33$ & HM \\
3C~33.1      & 0.1810 &  0.86 & 1.13 &  15.0  &  $-1.83$ & vBJ \\
3C~35        & 0.0670 &  0.59 & 1.31 &  14.0  &  $-1.65$ & vBJ \\
3C~42        & 0.3950 &  0.84 & 1.03 &   2.4  &  $-2.69$ & FBP \\
3C~47        & 0.4250 &  1.10 & 1.01 &  80.0  &  $-1.27$ & PH \\
3C~61.1      & 0.1860 &  1.91 & 1.08 &   2.3  &  $-3.00$ & GFGP \\
3C~67        & 0.3012 &  0.91 & 0.99 & 337.0  &  $-0.40$ & vBMH \\
3C~79        & 0.2559 &  1.31 & 1.01 &  10.5  &  $-2.19$ & SMP \\
3C~98        & 0.0306 &  4.97 & 0.96 &   9.0  &  $-2.75$ & FB \\
3C~109       & 0.3056 &  1.64 & 0.64 & 320.0  &  $-0.71$ & RP \\
4C~14.11$^*$ & 0.2070 &  0.89 & 0.73 &  29.7  &  $-1.52$ & HAPR \\
3C~123       & 0.2177 & 16.32 & 0.83 & 108.9  &  $-2.24$ & HAPR \\
3C~132       & 0.2140 &  1.05 & 1.00 &   4.1  &  $-2.49$ & HAPR \\
3C~153       & 0.2769 &  1.35 & 0.87 & $<0.5$ & $<-3.52$ & HAPR \\
3C~171       & 0.2384 &  1.22 & 0.80 &   2.2  &  $-2.82$ & GFGP \\
3C~173.1     & 0.2920 &  0.77 & 1.01 &   7.4  &  $-2.13$ & GFGP \\
3C~184.1     & 0.1182 &  1.24 & 0.73 &   6.0  &  $-2.35$ & RP \\
DA~240$^*$   & 0.0350 &  0.69 & 0.95 & 121.9  &  $-0.69$ & MKOW \\
3C~192       & 0.0598 &  2.05 & 0.71 &   6.0  &  $-2.55$ & H79 \\
4C~14.27$^*$ & 0.3920 &  0.39 & 0.99 & $<0.5$ & $<-3.03$ & LP \\
3C~215       & 0.4110 &  0.41 & 0.85 &  20.0  &  $-1.42$ & PH \\
3C~219       & 0.1744 &  2.29 & 1.04 &  51.6  &  $-1.71$ & HAPR \\
3C~223       & 0.1368 &  1.29 & 0.75 &   9.0  &  $-2.20$ & RP \\
4C~73.08$^*$ & 0.0581 &  0.90 & 0.84 &  19.0  &  $-1.69$ & M79 \\
3C~234       & 0.1848 &  1.54 & 1.04 &  90.0  &  $-1.29$ & RP \\
3C~244.1     & 0.4280 &  1.12 & 0.89 & $<0.6$ & $<-3.41$ & GFGP \\
3C~249.1     & 0.3110 &  0.78 & 0.94 & 110.0  &  $-0.91$ & PH \\
3C~268.3     & 0.3710 &  1.09 & 0.96 &$<20.0$ & $<-1.86$ & F90 \\
3C~274.1     & 0.4220 &  0.76 & 1.08 & $<6.0$ & $<-2.27$ & JPR \\
3C~284       & 0.2394 &  0.69 & 0.71 &   3.2  &  $-2.40$ & GFGP \\
3C~285       & 0.0794 &  0.76 & 0.81 &   6.8  &  $-2.07$ & HAPR \\
3C~300       & 0.2700 &  1.10 & 0.90 &   9.0  &  $-2.18$ & RP \\
3C~303       & 0.1410 &  0.94 & 0.80 & 150.0  &  $-0.78$ & PH \\
3C~319       & 0.1920 &  0.65 & 1.07 & $<1.0$ & $<-2.89$ & BF \\
3C~321       & 0.0960 &  1.22 & 0.83 &  30.0  &  $-1.63$ & JPR \\
3C~326       & 0.0895 &  0.48 & 1.63 &   3.5  &  $-2.20$ & R90 \\
3C~349       & 0.2050 &  1.14 & 0.80 &  24.0  &  $-1.73$ & JPR \\
3C~351       & 0.3710 &  1.21 & 0.84 &  15.0  &  $-2.02$ & L81 \\
3C~381       & 0.1605 &  1.29 & 0.94 &   5.2  &  $-2.45$ & F84 \\
3C~382       & 0.0578 &  2.22 & 0.74 & 251.2  &  $-0.91$ & HAPR \\
3C~388       & 0.0908 &  1.77 & 0.91 &  57.9  &  $-1.51$ & HAPR \\
3C~390.3     & 0.0569 &  4.48 & 0.62 & 340.0  &  $-1.10$ & vBJ \\
3C~401       & 0.2010 &  1.37 & 1.14 &  32.0  &  $-1.71$ & BBDG \\
3C~436       & 0.2145 &  0.99 & 0.99 &  20.0  &  $-1.77$ & RP \\
3C~438       & 0.2900 &  1.54 & 1.20 &  10.0  &  $-2.32$ & L81 \\
3C~452       & 0.0811 &  3.26 & 0.95 & 130.0  &  $-1.42$ & RP \\
3C~457       & 0.4270 &  0.55 & 0.93 &   6.0  &  $-2.10$ & LRL \\
\hline
\end{tabular}
\end{center}
\end{table}

The first step in our procedure is to determine the distribution of
the transverse core dominance parameter, $R_{\rm T}$.  We construct
synthetic samples of 5000 radio sources which are randomly oriented
with respect to the line of sight, and have the logarithms of their
transverse core dominance parameters, $\log R_{\rm T}$, drawn from a
Normal distribution, whose mean and standard deviation we vary to
determine the best fit to the observed data. We quantify the
goodness-of-fit using the Peto-Prentice generalized Wilcoxon test
(Latta 1981; Feigelson \& Nelson 1985), as employed in the {\sc asurv}
package within {\sc iraf/stsdas}. Of the various tests Feigelson \&
Nelson list to compare two samples of censored data, this is believed
to be the least affected by differences in sample size and censoring
patterns, which is important here since our synthetic sample is much
larger than the real dataset and suffers no censoring. The best fit we
find is when the transverse core dominance parameter is drawn from the
distribution ${\sf N}(-2.54,0.51)$, with the probability that the two
distributions are drawn from the same population being 99.9\%.  Both
versions of the Gehan test (also available in {\sc asurv}) produce the
same result, so it is not sensitive to the specific test used. Note
that the underlying distribution, $\log R_{\rm T} \in {\sf
N}(-2.43,0.50)$, reported by Simpson (1998b) was computed using a few
older core flux measurements (generally upper limits which have been
superseded by more sensitive limits or faint detections), and with a
different goodness-of-fit statistic. However, there is only a slight
difference between the two distributions, which does not affect the
results in that paper.

Having derived a distribution function for $\log R_{\rm T}$, it
becomes possible to infer the probability distribution of viewing
angles, $P(\theta|R)$. Bayes' Theorem informs us that
\[
P(\theta | R) \propto P(R | \theta) P(\theta) .
\]
Since $R$ and $\theta$ are directly related by $R_{\rm T}$, we can
write the first term on the right hand side as an {\em a priori\/}
probability that $R_{\rm T}$ lies in the appropriate range to produce
the observed value of $R$ given that the viewing angle lies in the
interval between $\theta$ and $\theta + \delta\theta$. If we define $x
\equiv \beta \cos \theta$, it is trivial to show that $R/R_{\rm T} =
(1+x^2) (1-x^2)^{-2}$. Hence,
\[
P(R | \theta) = P (\log R_{\rm T} = \log R \frac{(1-x^2)^2}{1+x^2})
\frac{{\rm d}\log R_{\rm T}}{{\rm d}\theta} ,
\]
and, straightforwardly,
\[
\frac{{\rm d}\,(\log R_{\rm T})}{{\rm d}\theta} = \frac{\beta \sin
\theta}{\ln 10} \frac{2x(3+x^2)}{1-x^4} .
\]

\begin{figure}
\includegraphics{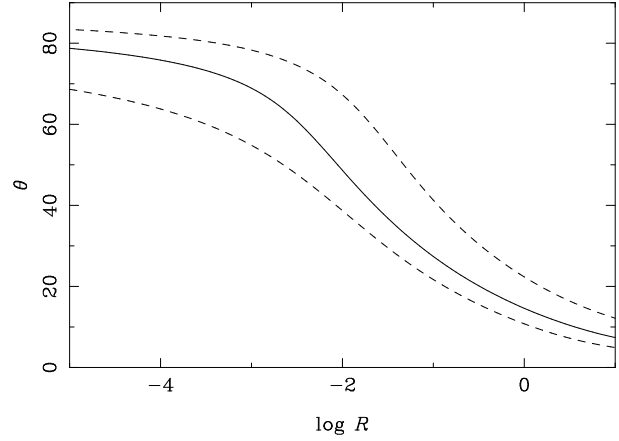}
\vspace*{64mm}
\caption[]{The {\em a posteriori\/} probability density function of
viewing angle, $\theta$, as a function of core-to-lobe ratio, $R$,
under the assumption described in the text. The solid line indicates
the maximum of the pdf, and the dashed lines show the $1\sigma$
limits.}
\label{fig:thetar}
\end{figure}

Finally, we need to consider the {\em a priori\/} distribution of
viewing angles, $P(\theta)$. A natural choice, motivated by the radio
galaxy--quasar unification scenarios, would be to assume that
$P(\theta) \propto \sin \theta$ for angles greater than some critical
angle, $\theta_{\rm c}$, and zero for angles less than this. In
practice, the actual choice of $\theta_{\rm c}$ has little effect on
the results, except for sources with bright radio cores, such as
3C~234, when the probability distribution of angles piles up at the
critical angle. Rather than adopt an {\em ad hoc\/} formulation to
create a smooth transition at $\theta \approx \theta_{\rm c}$ and
prevent this unrealistic behaviour, we choose $\theta_{\rm c} = 0$.
The behaviour of $\theta(R)$ under these assumptions is shown
graphically in Fig.~\ref{fig:thetar}.

\end{document}